\documentclass{PoS}
\usepackage{epsfig}
\usepackage{amsmath}
\usepackage{amssymb}
\usepackage{graphicx}
\def\be{\begin{equation}}
\def\ee{\end{equation}}
\def\bea{\begin{eqnarray*}}
\def\eea{\end{eqnarray*}}

\def\gsim{\lower0.5ex\hbox{$\:\buildrel >\over\sim\:$}} 

\def\lsim{\lower0.5ex\hbox{$\:\buildrel <\over\sim\:$}}

\title{Neutrinoless double $\beta$ decays
of the top quark and other effects of heavy Majorana neutrinos}
 
\ShortTitle{Neutrinoless double $\beta$ decays of the top quark
and other effects of heavy Majorana neutrinos}

\author{Gad Eilam\\
Technion - Isreal Institute of Technology, Haifa, Israel\\ 
Email: \email{eilam@physics.technion.ac.il}}   
 
\abstract{We discuss the rare decay of the top quark into a pair
of same charge leptons (with identical or different flavors), a $b$ quark
and a (real or virtual) $W^-$. The above process proceeds only if the 
exchanged neutrino $N$ is 
of the Majorana type. This decay is the neutrinoless
double $\beta$ decay of the top. We find measurable values  
for its rate at the LHC with luminosity of $100\rm{fb}^{-1}$.
Furthermore, we consider an interaction of charged Higgs bosons
with $N$      
which leads to lepton number violating processes
such as $pp\to \ell^+ N \to \ell^+ \ell^+ H^-$, exhibiting spectacular 
events of the type
$\ell^+\ell^+ b {\bar b}+2$ jets.}  

\FullConference{Physics at LHC 2008\\
                 29 September - October 4, 2008\\
                 Split, Croatia}

\begin{document}

The outline of this talk is as follows:\\
\indent I will first motivate the study of heavy Majorana neutrinos and top 
quarks. 
Subsequently, Majorana neutrinos
and lepton number violating (LNV) signals in $t$ quark and $W$ boson
rare decays will be discussed.
Then, I let $H^+$ and $H^-$ enter the game and briefly consider 
charged Higgs  
effects in the production and 
decay of a heavy Majorana neutrino at the Large Hadron Collider (LHC).\\
\indent Let me present the motivation to study heavy Majorana neutrinos 
and top quarks.
The discovery of neutrino oscillations implies that
they have mass~\cite{Kayser:2008rd}. Thus indicating that there is New
Physics (NP) beyond the Standard Model (SM).
A simple way to consistently include sub-eV neutrinos into the SM, 
is to add superheavy 
right-handed neutrinos with GUT-scale masses and to 
rely on 
the seesaw mechanism~\cite{Minkowski:1977sc} which yields a desired
light neutrinos mass scale: 
$m_\nu \sim M_{EW}^2/M_{GUT}
\sim 10^{-2}~{\rm eV}$.
This seesaw mechanism links neutrino masses
with NP at the GUT scale and raises the 
possibility that neutrinos will be of the Majorana 
type, where they are the antiparticles of themselves.
The mechanism through which neutrinos acquire mass is 
yet unknown and might be different from the seesaw one.
We take a purely phenomenological approach in which the masses 
of the heavy neutrinos are not predetermined by a specific 
model and will be taken here within the range explorable at
the LHC~\cite{NatLHC}.\\ 
\indent Most of us believe 
that something new is lurking over there, at about the
scale of EW symmetry breaking, which is around
250 GeV. The top quark at 172.6 GeV~\cite{topmass}   
is the quark closest to that scale and is therefore
the most sensitive to NP. For example,
various models of NP may lead
to Flavor Changing Neutral Currents (FCNC) of top, 
such as $t\to cH^0$~\cite{NPtcH}, which may be 10
orders of magnitude larger than their SM values~\cite{SMtcH}.\\
\indent The well known case which proceeds only if 
the light neutrino is Majorana,
and is LNV with $\Delta L=2$,       
is the neutrinoless double beta decay ($0\nu 2\beta$)~\cite{0nu2be}
$(A,Z)\to (A,Z+2) + e^- + e^-$ 
which has not been observed 
and it suffers from nuclear effects. Nevertheless, 
it produces very useful limits.
Also interesting are the
$\Delta L=2$, LNV processes 
in various high-energy collisions such as
$e^-~e^-\to W^-~W^-$ and in rare charged meson and lepton 
decays~\cite{atre09}.\\
\indent We explored~\cite{our1st} two additional LNV decay channels of the
real top quark and of the real W boson, to like sign lepton pairs:
$t \to b \ell^+_i \ell^+_j W^-$, and
$W^+ \to \ell^+_i \ell^+_j f \bar f^\prime$.
The above top decay proceeds through the following
diagrams:

\begin{center}
\includegraphics[height=7.0cm,width=10.1cm,angle=0]{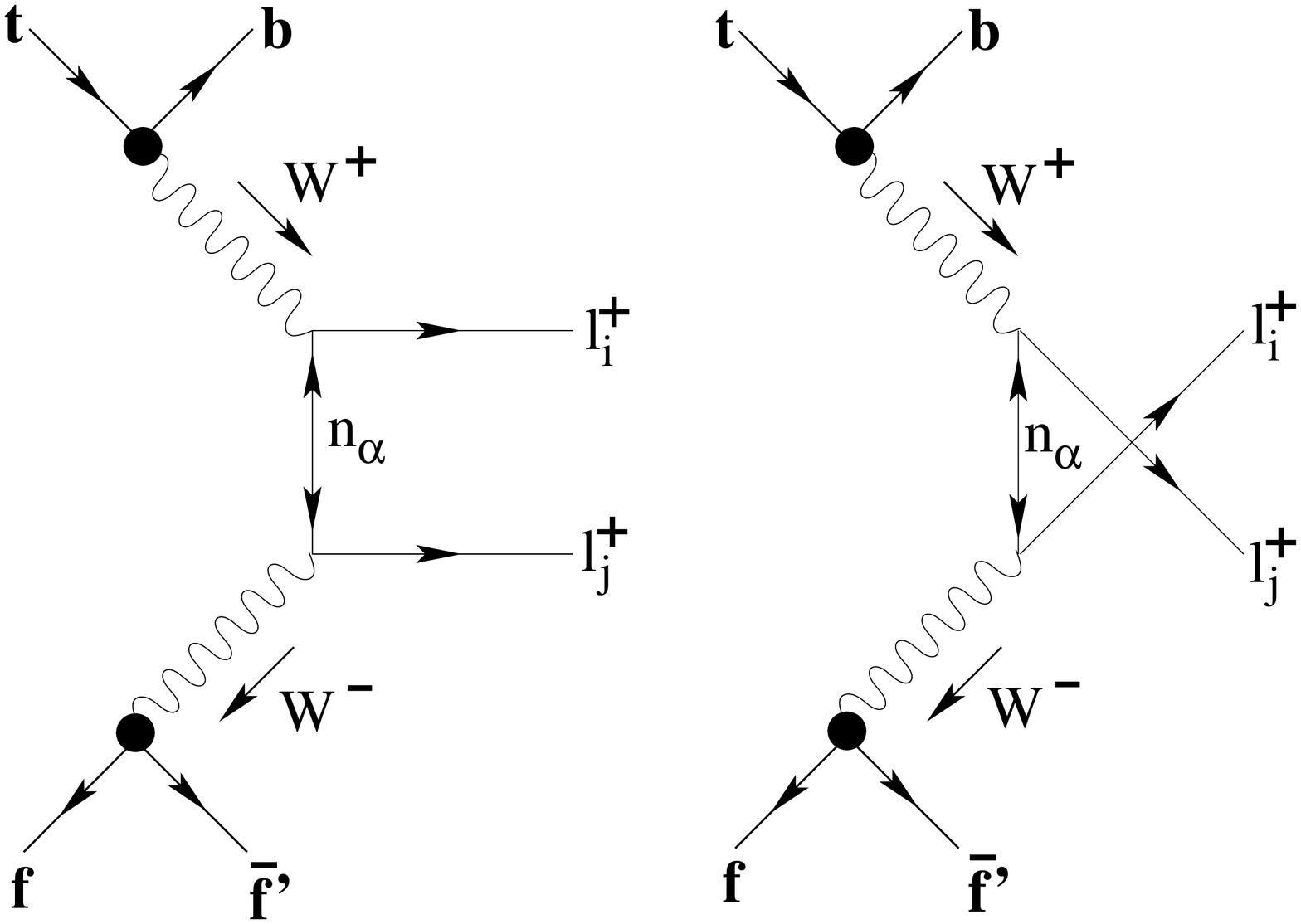}
\end{center}

These decays are induced by heavy Majorana
neutrino exchanges and may therefore serve as important tests of the
neutrino sector and as a possible evidence for the existence of
Majorana-type heavy neutrinos with masses at the EW scale.  
They both emanate from the same kernel:
$W^\pm W^\pm \to \ell_i^\pm \ell_j^\pm$,
with an
exchange of a Majorana neutrino. This is the same
kernel which induces $0\nu 2\beta$  in nuclei. However, in contrast to
the nuclear case, the $0\nu 2\beta$ decays of $t$ and $W$
are dominated by the exchanges of heavy (EW scale) neutrinos
instead of sub-eV neutrinos.\\
\indent For the $t$ case the top-quark decays to an 
on-shell $W$ boson with a ``wrong'' charge, 
as compared with the
"right" charge $W$ boson in the main decay $t\to b W^+$.
The above "weird" $t$ and $W$ decays, 
originate from:
\begin{eqnarray*}
{\cal L}=- \frac{g}{2 \sqrt{2}} B_{in}
W_\mu^- \ell_i \gamma^\mu (1- \gamma_5)
n_{\alpha}
+ H.c. 
\end{eqnarray*}
\noindent $\alpha=1-6$ stand for:
6 Majorana neutrino states, 3 of which are 
light and 3 are heavy.
$B$ is a $3 \times 6$ matrix with elements 
$B_{in} \equiv \sum_{k=1}^3 V_{ki}^L U_{kn}^*$ where
$V^L$ is the $3 \times 3$ unitary mixing matrix of the
left-handed charged leptons 
and $U$ is the $6 \times 6$ unitary mixing matrix in the neutrino sector.\\
\indent The possibility of non-seesaw realizations
or internal symmetries in the neutrino sector that decouple
the heavy-to-light neutrino mixing from 
neutrino masses, cannot
be excluded.
In a model independent
approach, the couplings
are bounded by 
experimental constraints.\\ 
\indent \underline{Assume:} a single  
heavy neutrino $N$ dominates.
The limits on its couplings to the charged 
leptons are expressed in terms
of $\Omega_{\ell \ell^\prime}
\equiv B_{\ell N} B_{\ell^\prime N}$.\\
\indent Limits on its flavor-diagonal 
couplings from precision electroweak data,
at 90\% CL, are~\cite{Kagan}:

\begin{eqnarray*}
\Omega_{ee} \leq 0.012 ~,~ \Omega_{\mu \mu} \leq 0.0096
~,~\Omega_{\tau \tau} \leq 0.016~,
\end{eqnarray*}

\noindent and on flavor-changing couplings, from limits
on rare flavor--violating lepton decays: 

\begin{eqnarray*} 
|\Omega_{e\mu}| \leq 0.0001 ~,~ |\Omega_{e \tau}| \leq 0.02 ~,~
|\Omega_{\mu \tau}| \leq 0.02 ~.
\end{eqnarray*}

The results of the calculations are depicted in the following 
figures:

\begin{center}
\includegraphics[height=7.0cm,width=10.5cm,angle=0]{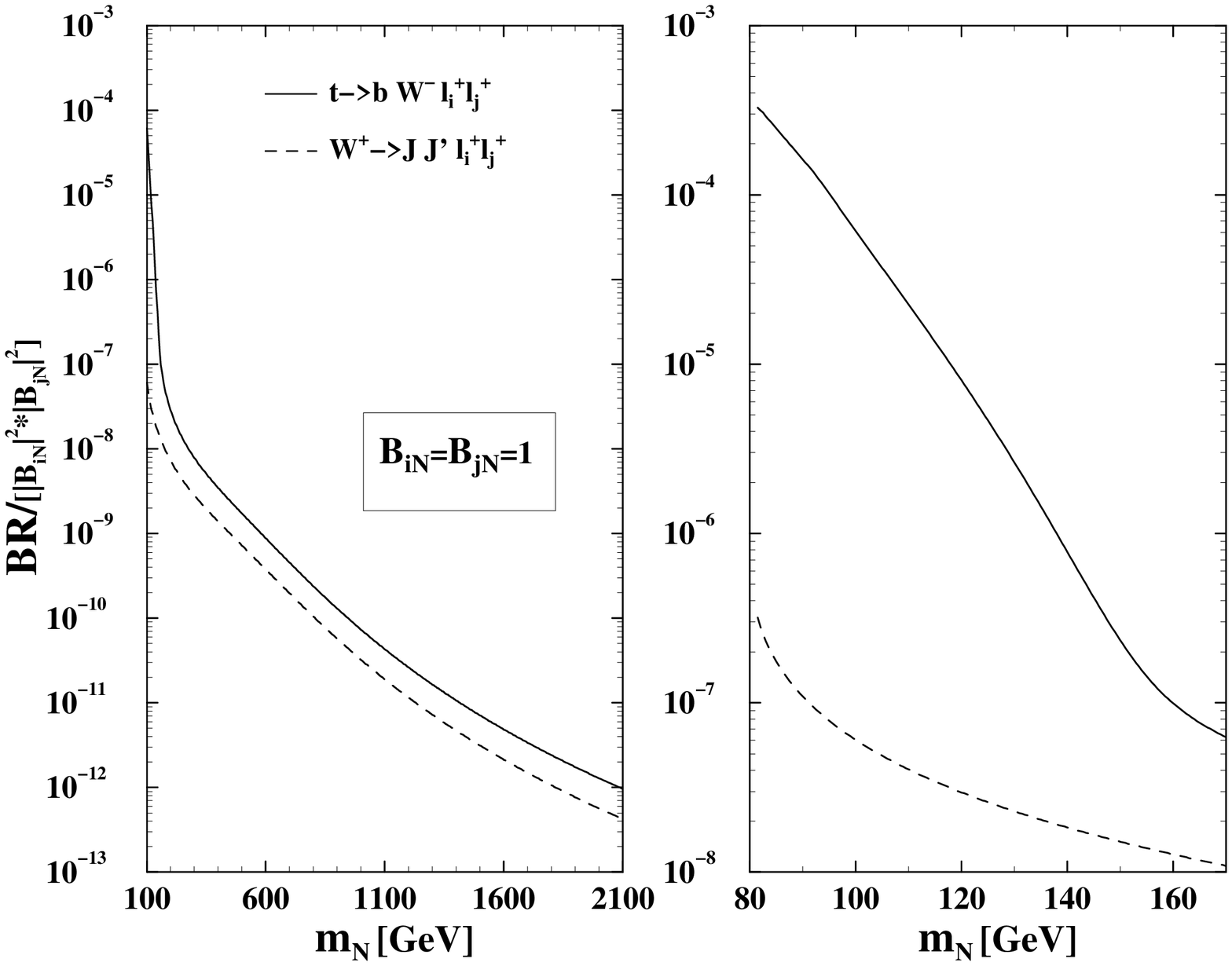}
\end{center}
  
The branching ratios (BRs) here are scaled 
by the mixing parameters i.e., we took
$B_{iN}=B_{jN}=1$, and are shown as a function of $m_N$, 
the Majorana neutrino mass. 
We see that, for both decays, a sizable 
BR can arise only
for $m_N \lsim 100$ GeV. Then: 
$BR(t \to b W^- \ell^+_i \ell^+)\sim 10^{-4}$ and 
$BR(W^+ \to J {\bar J}^\prime \ell^+_i \ell^+_j) \sim 10^{-7}$.\\
\indent For a more realistic BRs we use the bounds on the couplings.
In case of the $W$-boson decay,
the largest BR is
of order of $10^{-10}$. This is too small to be observed at the LHC, where
about $10^9 -10^{10}$ inclusive on-shell $W$'s are 
expected to be produced through
$p p \to W +X$, at an integrated luminosity of ${\cal O}(100)$
fb$^{-1}.$\\
\indent However, for more energetic off-shell $W$-bosons, produced at 
hadron colliders in the s-channel via $ud\to W^*$, the
sensitivity to the heavy Majorana exchanges can be significantly 
enhanced~\cite{Wstar}.
Indeed, it was demonstrated that the
s-channel $W^*$ can ``decay'' as
$W^{*} \to 2~{\rm jets}~+~\ell^\pm_i \ell^\pm_j$, by first decaying
to an on-shell Majorana neutrino $W^* \to \ell N$,
followed by $N \to \ell W \to \ell+~2~{\rm jets}$.
This process is easily accessible at the LHC.\\
\indent In the case of the top-quark decay  $t \to b W^- \ell^+_i \ell^+_j$,
taking $m_N \sim 100$ GeV and using the limits we discussed,
the BRs for the various $\ell^+_i \ell^+_j$ channels are given in the 
following table:

\begin{center}
\begin{tabular}{c||c|c|c|c|c|c}
\hline
&\multicolumn{6}{c}{$BR(t \to b W^- \ell^+_i \ell^+_j) \times 10^6$} \\
$\ell_i \ell_j=$ & $ee$ & $\mu \mu$ & $\tau \tau$ & $e \mu$ & $e \tau$ & $\mu \tau$ \\
\hline
$m_N =90$ GeV   & $1.4$ & $1.1$ & $1.9$  & $1.1 \cdot 10^{-4}$ &  $1.6$ & $1.4$ \\
$m_N =100$ GeV   & $0.6$  & $0.5$  & $0.8$ & $0.4\cdot 10^{-4}$ &  $0.7$ & $0.6$ \\
\hline
\end{tabular}
\end{center}

The cross-section for $t \bar t$ production at 14 TeV
is $\sim 850$ pb, yielding about $10^8$ $t \bar t$
pairs at an integrated luminosity of ${\cal O}(100)$ fb$^{-1}$.
Thus, a $BR(t \to b W^- \ell^+_i \ell^+_j) \sim 10^{-6}$ that can
arise in most $\ell^+_i \ell^+_j$ channels,
should be accessible at the LHC. In particular, the flavor-changing
decay channels $t \to b W^- e \tau$ and $t \to b W^- \mu \tau$ seem
to be the most promising ones (in spite of the low $\tau$ detection
efficiency) as the they are expected to be the cleanest with respect
to background.\\
\indent Note that a Majorana exchange is not necessarily the only mechanism
leading to $\Delta L=2$ processes. One can envisage, for instance, a
situation in which another type of new physics contributes together
with the Majorana exchange. Viable examples are R-parity
violating supersymmetry, or leptoquark exchange.
In the (rather contrived) cases like these it is in principle possible to
obtain destructive interference between the different mechanisms,
thus evading the limits on the couplings
considered here. Therefore, the rather sizable branching ratios,
obtained for ${\cal O}(1)$ mixing angles cannot be
excluded.\\
\indent Now we change gear and let $H^+$ and $H^-$ enter the game,
discussing charged Higgs-boson 
effects in the production and 
decay of a heavy Majorana neutrino at the LHC~\cite{our2nd}.\\
\indent We considered a new  interaction  between a
heavy Majorana neutrino and a charged Higgs boson,

\begin{eqnarray*}
{\cal L}=\frac{g}{2 \sqrt{2}}\xi_{\ell N} \frac{m_N}{m_W} {\bar N}
 (1- \gamma_5)\ell H^+ + H.c. 
\end{eqnarray*}

\noindent where $\xi_{\ell N}$ are dimensionless parameters 
with sizes determined by the underlying NP.\\
\indent We showed that ${\cal L}$  can have drastic implications on 
LNV signals with same-sign dileptons at the LHC.
The  LNV signal of heavy Majorana neutrinos previously considered
at the LHC, $pp \to \ell^+ N \to \ell^+ \ell^+ W^-$,
may be overwhelmed by $pp \to \ell^+ N \to \ell^+ \ell^+ H^-$.
With the subsequent decays $H^- \to \bar t b$ or $H^- \to W^- H^0$, 
the heavy Majorana
neutrino production leads to the spectacular events 
$\ell^+ \ell^+\ b \bar  b~+$ 2 jets.\\
\indent We also explored the case $m_N < m_{H^+}$,
where the decay $H^+ \to \ell^+ N$ can become the dominant
$N$-production mechanism at the LHC. In particular, we show that the process
$g\bar b \to \bar t H^+$ followed by $t \to \bar bW^-$ and 
$H^+ \to \ell^+ N \to \ell^+ \ell^+ W^-$ could lead to another type 
of spectacular events of $\ell^+ \ell^+\ b~+$ 4 jets.\\
\indent Now, a sentence to end with: Let us hope the LHC will be all in   
one i.e.,
that it will act both as a Higgs and as a  
top factory and that it will teach
us about lepton number violation and ...\\

\noindent {\bf Acknowledgment}

I would like to thank my colleagues, especially Shaouly Bar-Shalom, for 
sharing their wisdom with me.


\begin{thebibliography}{99} 
   \bibitem{Kayser:2008rd}
For a recent review, see:  B.~Kayser,
 \emph{Neutrino mass, mixing, and flavor change}, p. 163 in: Particle
Data Group (A. Amsler et al.), 
\emph{Review of Particle Physics},
Phys. Lett. {\bf B667} (2008) 1 
{\tt 0804.1497[hep-ph]} and references therein.
\bibitem{Minkowski:1977sc}
P.~Minkowski,
\emph{$\mu \to e \gamma$ at a rate of one out of 1-billion muon decays?}
\emph{Phys. Lett.} {\bf B67} (1977) 421; 
T. Yanagida, 
\emph{Horizontal symmetry and masses of neutrinos}, in proceedings  
of \emph{Workshop on the unified theory and the baryon number in the  
universe}, edited by A. Sawada and A. Sugamoto,
KEK, Tsukuba, Japan, 1979, p. 95; 
M. Gell-Mann, P. Ramond and R. Slansky,
\emph{Complex spinors and unified theories}, in proceedings of
\emph{Supergravity}, edited by P. van 
Nieuwenhuizen and D.Z. Freedman, North Holland, 1979, p. 315;
R.N. Mohapatra G. and Senjanovic,
\emph{Neutrino mass and spontaneous parity nonconservation},
Phys. Rev. Lett. {\bf 44} (1980) 912.
\bibitem{NatLHC}
For an example of such a model see: A.~Pilaftsis,
\emph{Radiatively induced neutrino masses and large Higgs neutrino couplings in
the standard model with Majorana fields},
  Z.\ Phys.\  C {\bf 55} (1992) 275.
\bibitem{topmass}
For this most recent result see: F.~Garberson (CDF and D0 collaborations),
\emph{Top quark mass and cross section results from the Tevatron},
{\tt 0808.0273 [hep-ex]}.
\bibitem{NPtcH}
For a recent review see: J.~M.~Yang,
\emph{Probing new physics from top quark processes at LHC: A mini review},
Int.\ J.\ Mod.\ Phys.\  A {\bf 23} (2008) 3343
{\tt 0801.0210[shep-ph]} and references therein.
\bibitem{SMtcH}
G.~Eilam, J.~L.~Hewett and A.~Soni,
\emph{Rare decays of the top quark in the standard and two Higgs doublet
models},
Phys.\ Rev.\  D {\bf 44} (1991) 1473
[Erratum-ibid.\  D {\bf 59} (1999) 039901];
B.~Mele, S.~Petrarca and A.~Soddu,
\emph{A new evaluation of the $t\to c H$ decay width in the standard model},
Phys.\ Lett.\  B {\bf 435} (1998) 401
{\tt hep-ph/9805498}   .
\bibitem{0nu2be}  
For a recent review see: P.~Vogel,
\emph{Neutrino mass and neutrinoless double beta decay}, in
proceedings of
\emph{4th international workshop on neutrino 
oscillations In Venice: Ten years after the neutrino oscillations}
edited by M. Baldo Ceolin, Papergraf, Padova, 2008, p. 75 
{\tt 0807.1559 [hep-ph]} and references therein.
\bibitem{atre09}
For an extensive list of references for
these and other processes involving neutrinos at the EW scale,
see:
A.~Atre, T.~Han, S.~Pascoli and B.~Zhang,
\emph{The Search for Heavy Majorana Neutrinos},
{\tt 0901.3589 [hep-ph]}.
\bibitem{our1st}
S.~Bar-Shalom, N.~G.~Deshpande, G.~Eilam, J.~Jiang and A.~Soni,
\emph{Majorana neutrinos and lepton-number-violating signals in top-quark and
W-boson rare decays}
Phys.\ Lett.\  B {\bf 643} (2006) 342
{\tt hep-ph/0608309};
Recently, a calculation similar to ours appeared, leading to similar
conclusions:
Z.~Si and K.~Wang,
\emph{GeV Majorana neutrinos in top-quark decay at the LHC},
Phys.\ Rev.\  D {\bf 79} (2009) 014034
{\tt 0810.5266 [hep-ph]}.
\bibitem{Kagan}
S.~Bergmann and A.~Kagan,
\emph{Z-induced FCNC and their effects on neutrino oscillations},
Nucl.\ Phys.\  B {\bf 538} (1999) 368
{\tt hep-ph/9803305}.
\bibitem{Wstar}
F.~M.~L.~Almeida, Y.~D.~A.~Coutinho, J.~A.~Martins Simoes and M.~A.~B.~do Vale,
\emph{Signature for heavy Majorana neutrinos in hadronic collisions},
Phys.\ Rev.\  D {\bf 62} (2000) 075004
{\tt hep-ph/0002024};
T.~Han and B.~Zhang,
\emph{Signatures for Majorana neutrinos at hadron colliders},
Phys.\ Rev.\ Lett.\  {\bf 97} (2006) 171804
{\tt hep-ph/0604064}.
\bibitem{our2nd}
S.~Bar-Shalom, G.~Eilam, T.~Han and A.~Soni,
\emph{Charged Higgs boson effects in the production 
and decay of a heavy Majorana neutrino at the LHC},
Phys.\ Rev.\  D {\bf 77} (200 8) 115019
{\tt 0803.28 35 [hep-ph]}.

\end{thebibliography}
\end{document}